\documentclass[prb,twocolumn,superscriptaddress]{revtex4}

\usepackage{amsmath}
\usepackage{amssymb}
\usepackage{epsfig}
\usepackage{graphicx}
\usepackage{wasysym}
\usepackage{bbm}

\newcommand \be{\begin{equation}}
\newcommand \ee{\end{equation}}
\newcommand \bes{\begin{equation*}} 
\newcommand \ees{\end{equation*}}
\newcommand \bea{\begin{eqnarray}}
\newcommand \eea{\end{eqnarray}}
\newcommand \bsea{\begin{subequations}\begin{eqnarray}} 
\newcommand \esea{\end{eqnarray}\end{subequations}}
\newcommand \beas{\begin{eqnarray*}} 
\newcommand \eeas{\end{eqnarray*}}
\newcommand \bfg{\begin{figure}}
\newcommand \efg{\end{figure}}
\newcommand \bfgs{\begin{figure*}} 
\newcommand \efgs{\end{figure*}}
\newcommand \bwt{\begin{widetext}}
\newcommand \ewt{\end{widetext}}

\newcommand \idmat{\mathbbm{1}} 
\newcommand \nd{{\vphantom{\dagger}}} 
\newcommand \bra{\langle}
\newcommand \ket{\rangle}
\newcommand \ua{\uparrow}
\newcommand \da{\downarrow}

\newcommand \eq{Eq.~}
\newcommand \eqs{Eqs.~}
\newcommand \fig{Fig.~}
\newcommand \refe{Ref.~}
\newcommand \refes{Refs.~}
\newcommand \sect{section~}
\newcommand \ssect{subsection~}

\newcommand \im{i} 
\newcommand \V[1]{\boldsymbol{#1}} 

\newcommand \vsigma{\V{\sigma}}
\newcommand \vtau{\V{\tau}}
\newcommand \vk{\V{k}}
\newcommand \vecr{\V{r}}
\newcommand \ve{\V{e}}

\newcommand \dif{\mathrm{d}} 
\newcommand \cdif{D} 
\newcommand \trace{\mathrm{Tr}} 

\newcommand \treverse{\mathcal{T}} 
\newcommand \qi{\mathbbm{i}} 
\newcommand \qj{\mathbbm{j}} 
\newcommand \qk{\mathbbm{k}} 

\newcommand \vM{\V{M}} 
\newcommand \vb{\V{b}} 
\newcommand \vK{\V{K}} 
\newcommand \spin{S} 
\newcommand \vspin{\V{\spin}}
\newcommand \orbital{T} 
\newcommand \vorbital{\V{\orbital}}
\newcommand \director{n} 
\newcommand \vdirector{\V{\director}}
\newcommand \ONdirector{\mathcal{N}} 
\newcommand \vONdirector{\V{\ONdirector}}

\newcommand \nnbond{\mathcal{U}} 
\newcommand \ansatz{\mathcal{A}} 
\newcommand \onsiteansatz{\eta} 
\newcommand \onsiteONansatz{\mu'} 
\newcommand \bondamplitude{f} 
\newcommand \rotor{\mathcal{Z}} 
\newcommand \gappedrotor{\mathcal{W}} 
\newcommand \spinon{\mathcal{B}} 
\newcommand \bfield{\hat{\mathcal{B}}} 

\usepackage{color}

\begin{document}

\title{Two-orbital Schwinger Boson Representation of Spin-One: 
Application to a Non-abelian Spin Liquid with Quaternion Gauge Field}

\author{Fa Wang}
\affiliation{Department of Physics, Massachusetts Institute of Technology, 
Cambridge, Massachusetts 02139, USA}

\author{Cenke Xu}
\affiliation{Department of Physics, University of California, 
Santa Barbara, CA 93106, USA}

\date{\today}

\begin{abstract}
A non-abelian spin liquid in triangular lattice spin-1 systems
was recently formulated in the form of continuum field theory
[T. Grover, and T. Senthil, Phys. Rev. Lett. {\bf 107}, 077203 (2011); 
Cenke Xu, A.W.W. Ludwig, arXiv:1012.5671]. 
It has spin-1/2 bosonic spinons coupled to emergent quaternion gauge fields,
and can be obtained by quantum disordering a non-collinear spin nematic order 
hypothesized to describe NiGa$_2$S$_4$ 
[H. Tsunetsugu, and M. Arikawa, J. Phys. Soc. Jpn. {\bf 75}, 083701 (2006)], 
However a microscopic lattice description, 
{e.g.} the lattice spinon (mean-field) Hamiltonian and the spin wavefunction, 
has been missing, 
and it has been noted that 
the standard Schwinger boson or bosonic triplon representations
of spin-1 cannot describe this spin liquid. 
In this paper a two-orbital Schwinger boson representation for spin-1 systems 
is developed and used to construct a mean-field description of 
this quaternion spin liquid.
Projecting the mean-field state produces a prototype wavefunction, 
which is a superposition of close-packed AKLT loop configurations
with nontrivial amplitudes. 
This new formalism and related wavefunctions may be generalized to higher 
spin systems and can possibly produce spin liquid states with 
even richer emergent gauge structures.
\end{abstract}


\maketitle

\tableofcontents

Spin liquid states in more than one spatial dimensions  
were proposed more than three decades ago\cite{Anderson}.
They are ground states of Mott insulators 
with no spontaneous symmetry breaking,
thus beyond the symmetry breaking paradigm of phases.
Many parent Hamiltonians of 
spin liquids with\cite{Klein,Kivelson-Klein,Nussinov-Klein,YaoH-LeeDH,YaoH-FuL} 
and without\cite{Kitaev,YaoH-3-12} 
spin $SU(2)$ symmetry have been constructed. Extensive numerical studies on 
semi-realistic models have shown evidences of 
spin liquid ground states in quantum 
spin models on triangular\cite{triangular-MSE-ED,Sheng-triangular,
 Sorella-triangular,Sheng-triangular-MSE,triangular-Hubbard-spin-ED} 
and kagome lattices\cite{kagome-ED,Sheng-kagome,White-kagome}
and also in electronic Hubbard models\cite{triangular-Hubbard,Assaad}.
In the last decade several promising candidate materials have also emerged, 
a review of which is given in \refe\onlinecite{Balents-Review}.

One way to understand some of the spin liquid states is by disordering 
a spin $SU(2)$ symmetry breaking order without proliferating 
topological defects\cite{Read-Sachdev-large-N,Sachdev-Read-large-N,Chubukov-Sachdev-Senthil,Senthil}.
Low energy theory of such description usually contains gapped bosonic spinon 
and emergent gauge field,
and the phase transition from quantum disordered (spin liquid) state 
to ordered state is the condensation of the bosonic spinon\cite{Read-Sachdev-large-N,Sachdev-Read-large-N,Chubukov-Sachdev-Senthil,Senthil}. 
This approach is believed to work better in the deep Mott insulating limit, 
where a short-range quantum spin model is appropriate.
To get quantum spin liquid ground state, the conventional wisdom suggests 
that low spin value is important, and spin-1/2 is the best.
For spin-1/2 systems the only single-site $SU(2)$-breaking order parameter
is the local magnetic dipole vector $\bra\vspin\ket$ 
(which also breaks time-reversal symmetry). 
Long-range order of $\bra\vspin\ket$ is the most commonly used 
starting point of this bosonic spin liquid approach\cite{Read-Sachdev-large-N,Sachdev-Read-large-N,Chubukov-Sachdev-Senthil,Senthil}.

However spin-1 systems may have magnetic quadrupole order
(spin nematic order hereafter) that breaks spin $SU(2)$ symmetry with 
{\em zero} local dipole moments and {\em no} time-reserval symmetry breaking.
The spin nematic order parameter is the real symmetric traceless matrix
 $Q^{ab}=\bra (\spin^a \spin^b+\spin^b \spin^a)/2\ket-(2/3)\delta^{ab}$
($a,b=x,y,z$).  
In this paper only the uniaxial spin nematic order will be considered, 
which can be described by the ``director'' $\vdirector$ as 
$Q^{ab}=\director^a \director^b-(1/3)\delta^{ab}\vdirector^2$.
 
\begin{figure}
\includegraphics[scale=0.6]{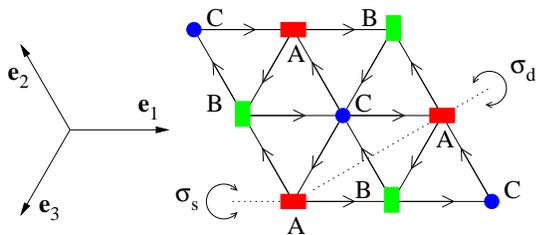}
\caption{(Color online)
Triangular lattice with the three sublattices labelled as $A,B,C$ respectively. 
$\ve_{1,2,3}$ are three lattice translation vectors.
$\sigma_s$ and $\sigma_d$ illustrate the two reflection generators of 
point group $C_{6v}$. 
An example of the three-sublattice spin 
nematic order\cite{Tsunetsugu} is illustrated, 
with the directors $\hat{x},\hat{y},\hat{z}$ drawn as
red horizontal bars, green vertical bars, and blue dots, respectively. 
The quaternion IGG elements on $A,B,C$ sites are the 
$A,B,C$ representations of $Q_8$ in \eqs(\ref{equ:classA}-\ref{equ:classC})
respectively. 
Arrow from site $r$ to $r'$ means the mean-field ansatz 
$\ansatz_{rr'}=-\ansatz_{r'r}^T=\nnbond=
(\idmat-\im\tau^x-\im\tau^y+\im\tau^z)/2$. 
}
\label{fig:lattice}
\end{figure}

Spin nematic orders have been proposed\cite{Tsunetsugu,Senthil-NiGaS} 
for the spin-1 triangular lattice material NiGa$_2$S$_4$ 
which had some experimental evidence of a ground state 
without magnetic dipole order\cite{NiGaS}.
In particular Tsunetsugu and Arikawa\cite{Tsunetsugu} proposed an interesting 
three-sublattice spin nematic order, with the directors on the 
three sublattices perpendicular to each other (see \fig\ref{fig:lattice}). 
This state was also found in a numerical study of spin-1 nearest-neighbor
bilinear-biquadratic Heisenberg model on triangular lattice\cite{Lauchli-ED}. 
Very recently two groups\cite{Grover-quaternion,XuCK-quaternion} 
considered possible spin liquid states 
by disordering this ``Tsunetsugu-Arikawa state''
(``antiferroquadrupolar order'' in \refe\onlinecite{Lauchli-ED}, 
``tetrad order'' in \refe\onlinecite{XuCK-quaternion}), 
and found an interesting non-abelian spin liquid 
with spin-1/2 bosonic spinons coupled to  
emergent gauge fields in the quaternion group $Q_8$, 
a discrete non-abelian group with eight elements defined as
$Q_8=\{\pm 1,\pm \qi,\pm \qj,\pm \qk\}$ 
with the multiplication rule $\qi^2=\qj^2=\qk^2=\qi\qj\qk=-1$. 
The continuum field theory for this spin liquid and for
the ordering transition, 
and many interesting properties, {e.g.} topological degeneracy, have been 
worked out in \refes\onlinecite{Grover-quaternion,XuCK-quaternion}. 
However no microscopic lattice spinon (mean-field) Hamiltonian was given 
and it was not clear how to construct a spin wavefunction 
for this ``quaternion spin liquid''.
The authors of \refe\onlinecite{Grover-quaternion} realized 
that it is impossible to describe this spin liquid 
by the Schwinger boson or ``bosonic triplon'' construction,
and suggested that ``such a spin liquid cannot be obtained by 
the standard projective construction for 
spin liquids''\cite{Grover-quaternion}. 
In this paper a new and non-standard projective construction 
in terms of a two-orbital Schwinger boson representation of spin-1
will be formulated, and used to construct a mean-field description of 
the quaternion spin liquid and its ordering transtion to the 
Tsunetsugu-Arikawa state, 
and produce a prototype spin wavefunction by projecting the mean-field state. 

Before proceeding to the new formalism 
it is worth reviewing how 
the old Schwinger boson and bosonic triplon constructions fail. 
The Schwinger boson representation for a spin-$S$ system\cite{Arovas-Auerbach}
is to write the spin operators as
($\vsigma$ are the spin Pauli matrices) 
\be
\vspin=\frac{1}{2}\sum_{s,s'=\ua,\da}
 b_{s}^\dagger \vsigma_{ss'}^\nd b_{s'}^\nd
,
\label{equ:SB}
\ee
with the constraint $\sum_{s} b_{s}^\dagger b_{s}^\nd = 2S$.
This representation has a $U(1)$ redundancy, $b\to e^{\im \theta}b$ 
($\im=\sqrt{-1}$, $\theta$ is real). 
For the gapped spin liquid phase to be stable in two-spatial-dimension(2D)
it is necessary to ``Higgs'' this $U(1)$ gauge field to a discrete subgroup, 
usually $Z_2$ by boson pairing\cite{Read-Sachdev-large-N,Sachdev-Read-large-N,Sachdev-kagome}. 
However the non-abelian quaternion gauge group is not a subgroup of this
abelian $U(1)$ gauge structure.
Moreover it is impossible to describe the spin nematic phase
in this formulation as a boson condensate, 
because a nonzero condensate $z=\bra b\ket$ necessarily produces 
a nonzero magnetic dipole moments 
$\vM=z^\dagger\cdot\vsigma^\nd\cdot z^\nd/2$ with size 
$|\vM|=|z|^2/2$. 

The bosonic triplon representation\cite{Papanicolaou,ZhangGM} 
uses a three-component boson $\vb$ 
and writes the spin operator as
\be
\vspin=-\im\vb^\dagger\times \vb^\nd
,
\ee
with the constraint $\vb^\dagger\cdot\vb=1$. 
Both magnetic order and spin nematic order can be described 
by a boson condensate in this construction. 
However the gauge freedom is again $U(1)$, $\vb\to e^{\im\theta}\vb$, 
with no quaternion subgroup. 
Moreover $\vb$ is not a spin-1/2 representation of spin $SU(2)$ symmetry. 
What is needed for the quaternion spin liquid is a new representation of spin-1 
by spin-1/2 bosons(like the Schwinger boson) with large enough gauge freedom, 
which can describe the spin nematic order semiclassically by 
boson condensation(like the bosonic triplon).

The outline of this paper is as follows. 
In \sect\ref{sec:rep} the two-orbital Schwinger boson representation
for spin-1 will be introduced.
Generic mean-field theories of spin liquids in this representation
and related gauge structure 
and generic (projected) mean-field wavefunctions will be presented.
In \sect\ref{sec:quaternion} a mean-field Hamiltonian
for the quaternion spin liquid on triangular lattice will be constructed 
and analyzed. 
The Tsunetsugu-Arikawa state will be obtained by boson condensation.
A prototype spin wavefunction for the spin liquid will be presented.
In \sect\ref{sec:discuss} remaining issues and possible extensions 
will be discussed.

\section{Two-orbital Schwinger Boson Representation of Spin-One 
and Mean-field Theories for Spin Liquids}
\label{sec:rep}

Spin-1 can be viewed as a symmetric combination of two spin-1/2.
The Affleck-Kennedy-Lieb-Tasaki(AKLT) state was originally
defined in this way\cite{AKLT}. 
Use this ``two-orbital'' picture and 
introduce Schwinger bosons for each spin-1/2, 
the spin-1 operators in terms of the four-component bosons are
\be
\vspin=\sum_{\alpha=1}^{2}\vspin_{\alpha}
=\sum_{\alpha=1}^{2}\frac{1}{2} \sum_{s,s'=\ua,\da}
 b_{\alpha s}^\dagger \vsigma_{ss'}^\nd b_{\alpha s'}^\nd
,
\label{equ:spin-1}
\ee
where $\alpha=1,2$ labels orbital, $s,s'=\ua,\da$ label spin. 
This bosonic representation for $SU(N)$ spins 
were briefly mentioned in \refe\onlinecite{Read-Sachdev-SUN}. 
From this alone the gauge freedom seems to be
orbital $U(2)$, namely 
$b_{\alpha s}\to \sum_{\beta} u_{\alpha\beta}b_{\beta s}$ with 
$\sum_{\beta} u_{\alpha\beta}^* u_{\alpha'\beta}^\nd=
\delta_{\alpha\alpha'}^\nd$. 
However the gauge transformations should also leave the constraints invariant. 

The first constraint requires that 
the total number of bosons is two,
\be
n=\sum_{\alpha} n_{\alpha}=
\sum_{\alpha s} b^\dagger_{\alpha s} b^\nd_{\alpha s}=2
.
\label{equ:constraint1}
\ee

Define the orbital pseudo-spins $\vorbital$ as
\be
\vorbital=\sum_{s,\alpha,\beta}\frac{1}{2} 
 b_{\alpha s}^\dagger \vtau_{\alpha\beta}^\nd b_{\beta s}^\nd
,
\label{equ:orbital-1}
\ee
where $\vtau$ are orbital Pauli matrices. 
There are ten states for two bosons, 
one $\spin=0,\orbital=0$ spin-orbital singlet
[$2^{-3/2}(b^\dagger\cdot\tau^y\sigma^y\cdot b^\dagger) |0\ket$
where $|0\ket$ is the boson vacuum]
and nine $\spin=1,\orbital=1$ states. 
The second constraint is to project out the singlet state,
\be
(b^\dagger\cdot\tau^y\sigma^y\cdot b^\dagger)
(b^\nd\cdot\tau^y\sigma^y\cdot b)=0
.
\label{equ:constraint2}
\ee

The $\spin=1,\orbital=1$ states can be arranged into a matrix, 
with row and column indices $\spin^z,\orbital^z=1,0,-1$, 
\be
\begin{pmatrix}
\frac{(b_{1\ua}^\dagger)^2}{\sqrt{2}}
&
b_{1\ua}^\dagger b_{2\ua}^\dagger
&
\frac{(b_{2\ua}^\dagger)^2}{\sqrt{2}}
\\
b_{1\ua}^\dagger b_{1\da}^\dagger
&
\frac{b_{1\ua}^\dagger b_{2\da}^\dagger+b_{1\da}^\dagger b_{2\ua}^\dagger}{\sqrt{2}}
&
b_{2\ua}^\dagger b_{2\da}^\dagger
\\
\frac{(b_{1\da}^\dagger)^2}{\sqrt{2}}
&
b_{1\da}^\dagger b_{2\da}^\dagger
&
\frac{(b_{2\da}^\dagger)^2}{\sqrt{2}}
\end{pmatrix}
|0\ket
.
\label{equ:S1T1-states}
\ee
The physical spin-1 states should be a linear combination of 
the columns in \eq(\ref{equ:S1T1-states}). 
This final constraint can be formally represented by
\be
\vONdirector\cdot\vorbital=0
,
\label{equ:constraint3}
\ee
with complex unit vector $\vONdirector$
($\vONdirector^*\cdot \vONdirector^\nd =1$). 
The chosen linear combination has coefficients
$-(\ONdirector^x-\im\ONdirector^y)/\sqrt{2},\ONdirector^z,
(\ONdirector^x+\im\ONdirector^y)/\sqrt{2}$
respectively for the three columns in \eq(\ref{equ:S1T1-states}).
An equivalent form of \eq(\ref{equ:constraint3}) is 
\be
\sum_{a,b=x,y,z}
\orbital^a\cdot \ONdirector^{a*} \ONdirector^b\cdot \orbital^b=0
.
\label{equ:constraint3-equivalent}
\ee
The gauge freedom should leave this ``vector Higgs condensate'' 
$\vONdirector$ invariant up to a complex phase, 
or equivalently leave the hermitian matrix $\ONdirector^{a*} \ONdirector^b$
invariant. 

This parton construction unifies the conventional Schwinger boson
\eq(\ref{equ:SB}) and the two-orbital AKLT representation. 
For example, 
$\vONdirector=(\hat{x}+\im\hat{y})/\sqrt{2}$ 
chooses the first column of \eq(\ref{equ:S1T1-states}) 
and is the old Schwinger boson representation. 
In contrast
$\vONdirector=\hat{z}$ chooses the middle column of \eq(\ref{equ:S1T1-states}) 
and is the AKLT representation.
Three different cases of $\vONdirector$ are discussed in the following. 

Case 1): $\vONdirector^*\times \vONdirector=0$, 
then a complex phase can be chosen so that $\vONdirector$ is a real vector. 
This represents an orbital nematic ordered state with 
the director $\vONdirector$, and is equivalent to the AKLT representation. 
The gauge freedom is $U(1)\times U(1)\rtimes Z_2$,
generated by
\be
b\to e^{\im\theta}b,\quad
b\to e^{\im\phi(\vONdirector\cdot\vtau)}b,\quad
b\to (\vONdirector_{\perp}\cdot\vtau)\cdot b
,
\label{equ:case1}
\ee
where $\vONdirector_{\perp}$ is a real unit vector perpendicular to
$\vONdirector$. 
Note that the last $Z_2$ does not commute with the second $U(1)$ group, 
therefore semidirect-product $\rtimes$ is used. 

Case 2): $\vONdirector^*\times \vONdirector\neq 0$ and
$\vONdirector\cdot \vONdirector=0$, then the real and imaginary parts
of $\vONdirector$ are perpendicular to each other and of equal length 
$\sqrt{1/2}$. 
This represents an orbital dipole ordered state with the orbital moment 
along $\hat{\vorbital}=-\im \vONdirector^*\times \vONdirector$, 
and is equivalent to the single-orbital 
Schwinger boson representation \eq(\ref{equ:SB}). 
The gauge freedom is 
$U(1)$, 
\be
b\to e^{\im \theta(\idmat+\hat{\vorbital}\cdot\vtau)} b
,
\label{equ:case2}
\ee
where $\idmat$ is the identity matrix.

Case 3): $\vONdirector^*\times \vONdirector\neq 0$ and
$\vONdirector\cdot \vONdirector\neq 0$, then 
a complex phase can be chosen so that
the real and imaginary parts
of $\vONdirector$ are perpendicular to each other but of different length. 
The orbital dipole moment is nonzero along $\hat{\vorbital} = 
-\im\vONdirector^*\times \vONdirector/|\vONdirector^*\times \vONdirector|$.
The gauge freedom is $U(1)\times Z_2$, 
generated by
\be
b\to e^{\im\theta}b,\quad
b\to (\hat{\vorbital}\cdot\vtau)\cdot b
.
\label{equ:case3}
\ee

Consider the Heisenberg antiferromagnetic interaction 
between two spin-1 at positions $r$ and $r'$, namely 
$\vspin_{r}\cdot\vspin_{r'}=\sum_{\alpha,\beta}\vspin_{r\alpha}\cdot\vspin_{r'\beta}$. 
The right-hand-side can be Hubbard-Stratonovich decoupled 
in the same way as the Schwinger boson mean-field theory\cite{Arovas-Auerbach}.
The mean-field Hamiltonian contains spin singlet boson pairing terms,
$\ansatz_{rr',\alpha\beta}^\nd
(b_{r,\alpha\ua}^\dagger b_{r',\beta\da}^\dagger
-b_{r,\alpha\da}^\dagger b_{r',\beta\ua}^\dagger)+{h.c.}$,
or in short form
$b_{r}^\dagger \cdot \ansatz_{rr'}^\nd\otimes 
 \im\sigma^y\cdot b_{r'}^\dagger+{h.c.}$
where $\ansatz_{rr'}$ is a generic $2\times 2$ matrix in the orbital space. 
Note that $\ansatz_{r'r}^{\vphantom{T}}=-\ansatz_{rr'}^T$ 
where superscript $^T$ stands for matrix transpose. 
For simplicity the boson hopping terms 
$\sum_{s} b^\dagger_{r,\alpha s}b^\nd_{r',\beta s}$ are
ignored. 

The constraints must be included in the mean-field theory 
by introducing onsite terms with Langrange multipliers\cite{WenXG-spinliquid}. 
The first constraint \eq(\ref{equ:constraint1}) can be incorporated by 
a chemical potential $\mu$ as $\mu (n-2)$. 
The second one \eq(\ref{equ:constraint2}) may be included as
$-\lambda (b^\dagger\cdot\tau^y\sigma^y\cdot b^\dagger)
(b^\nd\cdot\tau^y\sigma^y\cdot b)$
with real $\lambda$, 
however this is not quadratic in terms of bosons.
A non-rigorous Hubbard-Stratonovich decoupling 
can be performed to reduce this term to
$-\onsiteansatz(b^\dagger\cdot\tau^y\sigma^y\cdot b^\dagger)
-\onsiteansatz^*(b^\nd\cdot\tau^y\sigma^y\cdot b)+|\onsiteansatz|^2/\lambda$ 
with a complex Langrange multiplier $\onsiteansatz$. 
This procedure may be made rigorous regardless of the sign of $\lambda$
by the tricks of \refe\onlinecite{Flint}.
The final constraint \eq(\ref{equ:constraint3})
may be included as
$-\onsiteONansatz \vONdirector\cdot \vorbital+{h.c.}$
with a complex Lagrange multiplier $\onsiteONansatz$. 
In summary the generic mean-field Hamiltonian for spin liquids is
\be
\begin{split}
H_{\rm MF}=
\ &
-\sum_{r,r'}[b_{r}^\dagger \cdot 
 \ansatz_{rr'}^\nd\otimes \im\sigma^y\cdot b_{r'}^\dagger+{h.c.}]
+\sum_{r}\mu_{r}^\nd n_{r}^\nd
\\ &
-\sum_{r}[\onsiteansatz_r 
 (b^\dagger_r \cdot\tau^y\sigma^y\cdot b^\dagger_r)+{h.c.}]
\\ &
-\sum_{r}[ b_{r}^\dagger\cdot\Re(\onsiteONansatz_r\vONdirector_r)
\cdot\vtau\cdot b_{r}^\nd]
+{const.},
\end{split}
\label{equ:HMF}
\ee
where $\Re$ means real part.
The mean-field constraints are
\be
\begin{split}
&
\bra n_{r}\ket_{\rm MF}=2,\quad
\bra b^\dagger_r \cdot\tau^y\sigma^y\cdot b^\dagger_r\ket_{\rm MF}=0,\quad
\\ &
\bra b_{r}^\dagger\cdot(\vONdirector_r\cdot\vtau)\cdot b_{r}^\nd\ket_{\rm MF}=0
.
\end{split}
\ee

The mean-field Hamiltonian \eq(\ref{equ:HMF}) is not gauge invariant. 
Under a site-dependent gauge transformation $b_r\to G(r)\cdot b_r$, 
the mean-field ansatz $\{\mu_r,\onsiteansatz_r,\onsiteONansatz_r,\vONdirector_r,
\ansatz_{rr'}\}$ should transform as
\be
\begin{split}
&
\ansatz_{rr'}\to
G(r)\cdot \ansatz_{rr'}\cdot G^T(r')
,\quad
\onsiteansatz_r\to e^{2\im\theta}\onsiteansatz_r
,
\\ &
\mu_r+\Re(\onsiteONansatz_r\vONdirector_r)\cdot\vtau
\to
G(r)\cdot
[\mu_r+\Re(\onsiteONansatz_r\vONdirector_r)\cdot\vtau]
\cdot G^{-1}(r)
,
\end{split}
\label{equ:ansatz-under-gauge}
\ee
where $\theta$ defines the orbital-independent $U(1)$ subgroup
in \eqs(\ref{equ:case1}-\ref{equ:case3}). 
Gauge-invariant fluxes can be defined in analogy to
the Schwinger boson or $Sp(N)$ boson theory\cite{Tchernyshyov}.
The loop expansion for the mean-field ground state energy 
can also be performed,
and a ``flux expulsion'' argument for Heisenberg models 
may be raised as well\cite{Tchernyshyov}.

A realistic spin-1 Hamiltonian may contain the biquadratic interactions 
$(\vspin_r\cdot\vspin_{r'})^2$ and multiple-spin interactions, which cannot be
simply decoupled into quadratic terms of bosons. In this situation it is 
better to view the mean-field theory as a variational approach. 
The mean-field ground state after projection to physical spin-1 space 
can be used as a variational wavefunction. 
This viewpoint will be adopted throughout this paper, so
no self-consistent equation of $\ansatz$ will be solved,
and the overall scale of the ansatz does not matter. 

The mean-field ground state $|{\rm MF}\ket$ is generically 
\be
|{\rm MF}\ket=
\exp
\Big [
\frac{1}{2}
\sum_{r,r'} b^\dagger_r\cdot \bondamplitude_{rr'}^\nd \otimes \im \sigma^y
\cdot  b^\dagger_{r'}
\Big ]
\big |
0
\big\ket
,
\label{equ:MFWF}
\ee
where $\bondamplitude_{rr'}$ are $2\times 2$ matrices in the orbital space,
and have the same symmetry and gauge transformation rule as 
the mean-field ansatz $\ansatz_{rr'}$,
{e.g.} $\bondamplitude_{r'r}^\nd=-\bondamplitude_{rr'}^T$. 
$r=r'$ term with 
$\bondamplitude_{rr}^\nd=-\bondamplitude_{rr}^T\propto \tau^y$
is allowed but creates only onsite spin singlet
and will be projected out. 
Each 
$\bondamplitude_{rr',\alpha\beta}^\nd 
(b^\dagger_{\alpha\ua}b^\dagger_{\beta\da}
-b^\dagger_{\alpha\da}b^\dagger_{\beta\ua})$ 
term creates a spin singlet from two spin-1/2 on bond $rr'$. 
The projection onto physical spin-1 space requires 
two bosons on every site  
and the onsite symmetrization of the two orbitals. 
The projected wave function $\mathcal{P}|{\rm MF}\ket$ is therefore
a superposition of close-packed (every site is covered once) 
loop configurations $\{\ell\}$,
and on each loop $\ell$ the spin-1 form an AKLT state, 
\be
\mathcal{P}|{\rm MF}\ket
=\sum_{\{\ell\}}
\prod_{\ell} W_{\ell} |{\rm AKLT\ on\ }\ell\ket 
.
\label{equ:PMFWF}
\ee
The ``close-packed'' loop configurations may involve bonds 
longer than nearest-neighbor.
The amplitude factor $W_{\ell}$ for a length-$L$ 
loop $\ell=(r_1 r_2\dots r_L)$ is 
\be
\begin{split}
W_{\ell}=
\ &
(3/4)^{L/2}\cdot N_L\cdot
\trace[\tau_{r_1} \bondamplitude_{r_1 r_2} \tau_{r_2} \bondamplitude_{r_2 r_3} 
\dots \tau_{r_L} \bondamplitude_{r_L r_1}]
,
\end{split}
\label{equ:loopweight}
\ee
where
$\trace$ means matrix trace, 
$N_L=\sqrt{1+3\cdot(-3)^{-L}}$, and
$\tau_r=-\im\tau^y\cdot(\vONdirector_r^*\cdot\vtau)$ 
comes from the contraint \eq(\ref{equ:constraint3}) 
($\tau_r=\tau^x$ when $\vONdirector_r=\hat{z}$).
The factor $(3/4)^{L/2}\cdot N_L$ is the overlap between 
one spin-1/2 dimer pattern and the AKLT state\cite{AKLT}. 
$(3/4)^{L/2}$ produces an overall factor for the wavefunction 
and can be omitted. 
$N_L\sim 1$ when $L$ is large. 
$W_{\ell}$ is gauge invariant up to a global factor,
due to the fact that 
$G^T(r)\cdot (-\im\tau^y)\cdot(\vONdirector_r^*\cdot\vtau)\cdot G(r)
=e^{2\im\theta}(-\im\tau^y)\cdot G^{-1}(r)\cdot (\vONdirector_r^*\cdot\vtau) \cdot G(r)
\sim (-\im\tau^y)\cdot (\vONdirector_r^*\cdot\vtau)$ 
up to a complex phase, for any $G(r)$ in the gauge group 
\eqs(\ref{equ:case1}-\ref{equ:case3}).
$L$ can be $2$ in which case the AKLT state is 
the ``double-bond'' spin singlet state formed by two spin-1.

Wavefunctions for spin-1/2 spinon and gauge flux excitations can be 
constructed as well. 
The mean-field state with two spinons at $r,r'$ is given by
\be
|\spinon_{r\vphantom{r'}}^\nd,\spinon_{r'}^\nd\ket_{\rm MF} =
\trace[\bfield_{r\vphantom{r'}}^\dagger\cdot\spinon_{r\vphantom{r'}}^\nd]
\trace[\bfield_{r'}^\dagger\cdot\spinon_{r'}^\nd]|{\rm MF}\ket
\ee
where $\spinon_{r,r'}$ are $2\times 2$ complex spinon state matrices, and
\be
\bfield=\begin{pmatrix} 
b_{1\ua} & b_{2\ua}\\
b_{1\da} & b_{2\da}
\end{pmatrix}
.
\label{equ:define-bfield}
\ee
Projecting this state onto spin-1 space creates a superposition
of configurations with one open AKLT chain $\ell_{rr'}$ from $r$ to $r'$ plus  
close-packed AKLT loops.
The direction of the end-spins of open AKLT chain are given by
$\trace[\spinon^\dagger\vsigma\spinon]$ 
(normalization requires $|\trace[\spinon^\dagger\vsigma\spinon]|=1$).
The amplitude for a length-$(L+2)$ open chain $\ell_{rr'}=(r r_1\dots r_L r')$ 
with the end-spins at $r,r'$ in $\spin^z$ eigenstates $s,s'=\ua,\da$ is
($L$ can be zero)
\be
\begin{split}
W_{\ell_{rr'},ss'}=
\ &
(3/4)^{L/2+1}\cdot N_{L,ss'}
\\ &
\times
(\spinon_r^\nd\tau_{r}\bondamplitude_{rr_1}^\nd
\tau_{r_1}\bondamplitude_{r_1r_2}^\nd\dots
\tau_{r_L}\bondamplitude_{r_L r'}^\nd\tau_{r'}\spinon_{r'}^T)_{ss'}^\nd
,
\end{split}
\label{equ:chainweight}
\ee
with\cite{AKLT} $N_{L,ss'}=\sqrt{1+(-3)^{-L-2} (\sigma^x-\idmat)_{ss'}}$.
Multiple (even) number of spinons can be constructed similarly. 
Gauge flux excitations and topological degeneracy of ground states
will be demonstrated in the quantum limit of quaternion spin liquid
in \ssect\ref{ssec:prototype}. 

Spin liquids described in this way will have gapped spin-1/2 bosonic
spinons. For them to be stable in 2D it is necessary to ``Higgs'' the
continuous compact gauge groups \eqs(\ref{equ:case1}-\ref{equ:case3}) to 
a discrete subgroup. 
Many possibilities exist which can in principle be completely classified by 
the projective symmetry group(PSG) method\cite{WenXG-PSG,WenXG-spinliquid,WenXG-ZhouY-PSG,Wang-SBPSG}. 
This brute-force approach will not be attempted here, 
but the PSG language will be used to show that the mean-field theory
indeed describes a spin liquid state with no symmetry breaking.
This will be achieved by the explicit construction of the PSG elements,
$b_r\to G_X(r)\cdot b_{X(r)}$, for all generators $X$ of the physical symmetry 
group (space group and time-reversal). The mean-field Hamiltonian 
shall be invariant under PSG actions. 

The case with uniform $\vONdirector_r=\hat{z}$ will be considered 
hereafter only, except \ssect\ref{ssec:different}. 
The physical spin-1 states $|\spin^z=+1,0,-1\ket$ are 
\bsea
|\spin^z=+1\ket & = & b_{1\ua}^\dagger b_{2\ua}^\dagger |0\ket
,\label{equ:state1}
\\
|\spin^z=0\ket & = & \frac{1}{\sqrt{2}} 
(b_{1\ua}^\dagger b_{2\da}^\dagger+b_{1\da}^\dagger b_{2\ua}^\dagger)|0\ket
,\label{equ:state2}
\\
|\spin^z=-1\ket & = & b_{1\da}^\dagger b_{2\da}^\dagger |0\ket
.
\label{equ:state3}
\esea
The constraints \eq(\ref{equ:constraint1}) and \eq(\ref{equ:constraint3})
can be rewritten as
\be
n_{\alpha}=\sum_{s} b^\dagger_{\alpha s} b^\nd_{\alpha s}=1,\quad \alpha=1,2
.
\label{equ:constraint13-z}
\ee
The $U(1)\times U(1)\rtimes Z_2$ gauge group is,
\be
\begin{split}
&
b\to
e^{\im \theta} (\cos\phi\cdot\idmat+\sin\phi\cdot\im\tau^z)\cdot b,\quad
{\rm or\ }
\\ &
b\to
e^{\im \theta} (\cos\phi\cdot \im\tau^x+\sin\phi\cdot \im\tau^y)\cdot b
. 
\end{split}
\label{equ:U1U1Z2}
\ee
For frustrated({e.g.} triangular) lattices 
the orbital-independent $U(1)$ freedom $e^{\im \theta}$
will be removed by boson pairings in mean-field theory. 
The remaining $U(1)\rtimes Z_2$ 
[a subgroup of $SU(2)$ by setting $\theta=0$ in \eq(\ref{equ:U1U1Z2})] 
will be the starting point of discussions hereafter. 
It is non-abelian and contains the quaternion group.

A semiclassical picture of the uniaxial spin nematic order 
is that the two spin-1/2 have dipole moments antiparallel to each other
and along the direction of the director,
$\bra \vspin_1\ket=-\bra \vspin_2\ket \propto \vdirector$. 
This can be achieved by a single spin-orbital-entangled condensate, 
{e.g.} $\bra(b_{1\ua},b_{1\da},b_{2\ua},b_{2\da})\ket\propto (1,0,0,1)$
for $\vdirector\propto \hat{z}$.
However the quadrupole order parameter $Q^{ab}$
is naively zero because $\bra \spin^a\ket$ is zero.
This can be remedied by recognizing that
$Q^{ab}=\bra \sum_{\alpha,\beta}
(\spin^a_{\alpha} \spin^b_{\beta}+\spin^b_{\alpha} \spin^a_{\beta})\ket/2
-(2/3)\delta^{ab}
=\bra \spin^a_1 \spin^b_2+\spin^b_1 \spin^a_2 \ket-(1/6)\delta^{ab}$,
and the last expression is nontrivial in this condensate state
[although not in the form of $n^a n^b-(1/3)\delta^{ab}\vdirector^2$]. 
This can be further justified by projecting the coherent state
from this condensate $\exp[w\,(b_{1\ua}^\dagger+b_{2\da}^\dagger)]|0\ket$
onto the physical spin-1 states \eqs(\ref{equ:state1}-\ref{equ:state3}),
which gives the uniaxial spin nematic state $(w^2/\sqrt{2})|\spin^z=0\ket$ 
with director along $\hat{z}$ direction. 
In general the nematic director from a condensate $\bra b\ket$ is given by 
$\bra b\ket^* \cdot \tau^z\vsigma\cdot \bra b\ket$.

\section{Quaternion Spin Liquid on Triangular Lattice}
\label{sec:quaternion}

For a mean-field theory of quaternion spin liquid, 
the invariant gauge group(IGG)\cite{WenXG-PSG,WenXG-spinliquid} 
must be a representation of the quaternion group 
$Q_8=\{\pm 1,\pm\qi,\pm\qj,\pm\qk\}$, 
with eight distinct IGG elements 
$G_q(r)\in U(1)\rtimes Z_2$
such that: 
1)
the ansatz $\{\mu_r,\onsiteansatz_r,\onsiteONansatz_r,\vONdirector_r=\hat{z},
\ansatz_{rr'}\}$
are invariant under the actions \eq(\ref{equ:ansatz-under-gauge}) of $G_q(r)$ 
for any $q\in Q_8$, 
and no other element of \eq(\ref{equ:U1U1Z2}) can do the same; 
2) 
$G_q(r)$ is a representation of $Q_8$ for any site $r$. 

There are three distinct classes (labelled by $A,B,C$) of 
$Q_8$ representations on a single site. 
Representatives of each class are given below 
($G_{\pm 1}=\pm \idmat$ for all classes),
\bsea
A:&& 
G_{\pm \qi}=\mp \im \tau^x,\ 
G_{\pm \qj}=\mp \im \tau^y,\   
G_{\pm \qk}=\mp \im \tau^z;
\label{equ:classA}
\\
B:&&
G_{\pm \qi}=\mp \im \tau^y,\ 
G_{\pm \qj}=\mp \im \tau^z,\   
G_{\pm \qk}=\mp \im \tau^x;
\label{equ:classB}
\\ 
C:&&
G_{\pm \qi}=\mp \im \tau^z,\ 
G_{\pm \qj}=\mp \im \tau^x,\   
G_{\pm \qk}=\mp \im \tau^y.
\label{equ:classC}
\esea
Each class is generated by group conjugacy on its representative,
$G_q\to G\cdot G_q\cdot G^{-1}$ for $G\in U(1)\rtimes Z_2$. 
Therefore by site($r$)-dependent gauge transformations 
all $G_q(r)$ can be reduced to one of those in 
\eqs(\ref{equ:classA}-\ref{equ:classC}).

These $Q_8$ IGGs will constrain allowed ansatzs.
For the onsite terms, 
the $Q_8$ IGGs demand $\onsiteONansatz_{r}=0$
but put no constraint on $\onsiteansatz_r$. 
On the converse, however, $\onsiteONansatz_{r}=0$ and $\onsiteansatz_r$ 
do not reduce the $U(1)\rtimes Z_2$ freedom.

Consider a bond $rr'$ with nonzero $\ansatz_{rr'}$. 
The $Q_8$ IGGs demand 
$\ansatz_{rr'}=G_q(r)\cdot \ansatz_{rr'}\cdot G_q(r')^T$ for all $q\in Q_8$. 
There are three possibilities for $\ansatz_{rr'}$
depending on the representation choice combinations $(rr')$, 
\be
\ansatz_{rr'}\propto
\left \{
\begin{array}{ll}
\tau^y, & (rr')=(AA),(BB),(CC),\\
\nnbond, & (rr')=(AB),(BC),(CA),\\
\nnbond^T, & (rr')=(AC),(BA),(CB),
\end{array}
\right.
\ee
where the $SU(2)$ matrix 
$\nnbond = (\idmat-\im\tau^x-\im\tau^y+\im\tau^z)/2$
will appear frequently. 
Consider the converse problem, namely whether $\ansatz_{rr'}$
can ``Higgs'' the gauge freedom to $Q_8$. 
$\ansatz_{rr'}\propto \tau^y$ will not do this job, 
because all $G(r)=G(r')\in U(1)\rtimes Z_2$ 
will keep $\ansatz_{rr'}\propto \tau^y$ invariant.
The other two possibilities $\ansatz_{rr'}=\nnbond$ or $\nnbond^T$ 
will reduce the gauge freedom to $Q_8$ with the representation choices
given above. 
For instance consider
$G(r')=\cos\phi\cdot\idmat+\sin\phi\cdot\im\tau^z$ and $\ansatz_{rr'}=\nnbond$, 
the constraint solves for 
$G(r)=\cos\phi\cdot\idmat+\sin\phi\cdot\im\tau^y$
which can be a member of the $U(1)\rtimes Z_2$ group 
only if $\phi$ is a integral multiple of $\pi/2$, 
restricting $G(r)$ and $G(r')$ to be members of $Q_8$ representations. 

\subsection{Mean-field Theory on Triangular Lattice}
\label{ssec:MFT}

With the above general considerations a mean-field Hamiltonian of 
quaternion spin liquid can be constructed on the triangular lattice. 
Due to the three-sublattice structure it is natural to assign the
three $Q_8$ representations to the three corresponding sublattices. 
In this paper only the nearest-neighbor ansatz 
will be considered, with $\ansatz_{rr'}=\nnbond$ or $-\nnbond^T$
as shown in \fig\ref{fig:lattice}. 
Note that by the variational interpretation 
the overall scale of $\ansatz_{rr'}$ does not matter,
and the overall complex phase of $\ansatz_{rr'}$ can be removed by
a global orbital-independent $U(1)$ phase rotation of bosons. 
Translation symmetry further requires uniform 
$\mu_{r}=\mu$ and $\onsiteansatz_r=\onsiteansatz$. 

The mean-field Hamiltonian reads (up to a constant), 
\be
\begin{split}
H_{\rm MF}=
\ &
\sum_{\vecr} 
\Big [
\mu\,n_{\vecr}
-(\onsiteansatz\,b^\dagger_{\vecr}
\cdot\tau^y\sigma^y\cdot b^\dagger_{\vecr}+h.c.)
\\ &
\phantom{
\sum_{\vecr} 
\Big [
}
-\sum_{d=1}^{3}
(
b^\dagger_{\vecr} \cdot \nnbond\otimes\im\sigma^y \cdot b^\dagger_{\vecr+\ve_d}
+{h.c.}
)
\Big ]
.
\end{split}
\label{equ:HMF-triangular}
\ee

Physical symmetries are generated by 
two translations $T_{1,2}$ along $\ve_{1,2}$,
two reflections $\sigma_s$ and $\sigma_d$ (see \fig\ref{fig:lattice}),
and time-reversal $\treverse$. 
$T_{1,2}$ and $\sigma_s$ are trivial. 
$\sigma_d$ reverses all bond orientations in \fig\ref{fig:lattice}.
$\treverse$ changes the ansatz to their complex conjugate. 
Corresponding PSG elements are, 
\bsea
T_{1,2} & : & b_{\vecr}\to b_{\vecr+\ve_{1,2}}
,
\label{equ:PSGb1}
\\
\sigma_s & : & b_{\vecr}\to b_{\sigma_s(\vecr)}
,
\label{equ:PSGb2}
\\
\sigma_d & : & b_{\vecr}\to 
\sqrt{1/2}(\im\tau^x-\im\tau^y)\cdot b_{\sigma_d(\vecr)}
,
\label{equ:PSGb3}
\\
\treverse & : & b_{\vecr}\to \tau^y\sigma^y\cdot b_{\vecr}
.
\label{equ:PSGb4}
\esea
The $Q_8$ IGG is defined in \eqs(\ref{equ:classA}-\ref{equ:classC}) 
for $A,B,C$ sublattices respectively.
Time-reversal symmetry restricts $\onsiteansatz$ to be real. 
With this PSG constructed the mean-field Hamiltonian describes 
a ``symmetric spin liquid''\cite{WenXG-spinliquid}
with no broken symmetry. 

The mean-field Hamiltonian can be solved in the same way as 
the Schwinger boson mean-field theories\cite{Sachdev-kagome}.
Do the Fourier transform 
$b_{\vk}=N_{\rm site}^{-1/2}\sum_{\vecr} e^{-\im\vk\cdot\vecr} b_{\vecr}$
where $N_{\rm site}$ is the number of sites, 
and define 
$\Psi_{\vk}=(
b_{\vk,1\ua}^\nd,b_{\vk,2\ua}^\nd,b_{-\vk,1\da}^\dagger,b_{-\vk,2\da}^\dagger
)^T$, 
\eq(\ref{equ:HMF-triangular}) becomes
\be
H_{\rm MF}=
\sum_{\vk}
\Big [
\Psi_{\vk}^\dagger \cdot
\begin{pmatrix}
\mu\,\idmat & -P_{\vk} \\
-P_{\vk}^\dagger & \mu\,\idmat
\end{pmatrix}
\cdot \Psi_{\vk}^\nd
-2\mu
\Big ]
.
\label{equ:HMF-k}
\ee
where $P_{\vk}= \im(\idmat-\im\tau^x+\im\tau^z)\,\Im\xi_{\vk}
-\im\tau^y(\Re\xi_{\vk}+\onsiteansatz)$, 
$\Re$ and $\Im$ are real and imaginary parts, 
$\xi_{\vk}=
 e^{\im \vk\cdot \ve_1}+e^{\im \vk\cdot \ve_2}+e^{\im \vk\cdot \ve_3}$.
Do a singular value decomposition 
\be
P_{\vk}=U_{\vk}^\nd\cdot 
\begin{pmatrix}\rho_{1}(\vk) & 0 \\ 0 & \rho_{2}(\vk)\end{pmatrix}
\cdot V_{\vk}^\dagger
\ee
with $U(2)$ matrices $U_{\vk},V_{\vk}$ given by
\bea
U_{\vk} & = &
\begin{pmatrix}
-\sqrt{\frac{1}{3+\sqrt{3}}} & \sqrt{\frac{1}{3-\sqrt{3}}}\\
e^{\im\pi/4}\sqrt{\frac{1}{3-\sqrt{3}}} &
 e^{\im\pi/4}\sqrt{\frac{1}{3+\sqrt{3}}}
\end{pmatrix}
,
\\
V_{\vk} & = &
\begin{pmatrix}
e^{\im\pi/4}\sqrt{\frac{1}{3-\sqrt{3}}} &
 -e^{\im\pi/4}\sqrt{\frac{1}{3+\sqrt{3}}}\\
\sqrt{\frac{1}{3+\sqrt{3}}} & \sqrt{\frac{1}{3-\sqrt{3}}}
\end{pmatrix}
,
\eea
and real singular values 
$\rho_{1,2}(\vk)
=\sqrt{3}\,\Im\xi_{\vk}\pm (\Re\xi_{\vk}+\onsiteansatz)
.
$
Define ``Bogoliubov quasiparticles''
\be
\Phi_{\vk}
=
\begin{pmatrix}
\gamma_{\vk,1\ua}\\
\gamma_{\vk,2\ua}\\
\gamma_{-\vk,1\da}^\dagger\\
\gamma_{-\vk,2\da}^\dagger
\end{pmatrix}
=
\begin{pmatrix}
C_1 & 0 & S_1 & 0\\
0 & C_2 & 0 & S_2\\
S_1 & 0 & C_1 & 0\\
0 & S_2 & 0 & C_2
\end{pmatrix}
\begin{pmatrix}
U_{\vk}^\dagger & 0_{2\times 2}\\
0_{2\times 2} & V_{\vk}^\dagger
\end{pmatrix}
\Psi_{\vk}
,
\ee
where 
$C_{1,2}=\sqrt{ 1+\mu/E_{1,2}(\vk) }/\sqrt{2}$ 
and 
$S_{1,2}=-\rho_{1,2}(\vk)/E_{1,2}(\vk)/2C_{1,2}$, 
with the mean-field dispersions 
\be
E_{1,2}(\vk)=\sqrt{\mu^2-\rho_{1,2}^2 (\vk)}
.
\label{equ:dispersion}
\ee
\eq(\ref{equ:HMF-k}) is diagonalized by this $SU(2,2)$ Bogoliubov 
transformation,
\be
\begin{split}
H_{\rm MF}=
\ &
\sum_{\vk}
\big [
E_{1}(\vk)(\gamma_{\vk,1\ua}^\dagger \gamma_{\vk,1\ua}^\nd
+\gamma_{-\vk,1\da}^\nd \gamma_{-\vk,1\da}^\dagger)
\\ &
\quad
+
E_{2}(\vk)(\gamma_{\vk,2\ua}^\dagger \gamma_{\vk,2\ua}^\nd
+\gamma_{-\vk,2\da}^\nd \gamma_{-\vk,2\da}^\dagger)
-2\mu
\big ]
.
\end{split}
\ee
The mean-field ground state energy per site is
\be
E_{\rm MF}=N_{\rm site}^{-1}\sum_{\vk}[E_1(\vk)+E_2(\vk)-2\mu]
.
\ee

The mean-field ground state is annihilated by all $\gamma_{\vk,\alpha s}$,
and is 
\be
|{\rm MF}\ket =\exp
\Big [
\frac{1}{2}\sum_{\vk} 
b^\dagger_{\vk}
\cdot \bondamplitude_{\vk}\otimes \im\sigma^y\cdot
b^\dagger_{-\vk}
\Big ]
|0\ket
\ee
where $\bondamplitude_{\vk}$ is the Fourier transform of $\bondamplitude_{rr'}$, 
\be
\begin{split}
&
\bondamplitude_{\vk}=-U_{\vk}^\nd\cdot 
\begin{pmatrix}S_1/C_1 & 0\\0 & S_2/C_2\end{pmatrix}
\cdot V_{\vk}^\dagger
,
\\ &
\bondamplitude_{\vecr\vecr'}=N_{\rm site}^{-1} \sum_{\vk} 
e^{\im \vk\cdot(\vecr-\vecr')} \bondamplitude_{\vk}
,
\end{split}
\label{equ:bondamplitude}
\ee
and $|0\ket $ is the vacuum of $b$ bosons.

The mean-field constraints are 
\be
2=\bra n\ket_{\rm MF}=\frac{\partial E_{\rm MF}}{\partial\mu}
=
\int \widetilde{\dif^2\vk}
\big [
\frac{\mu}{E_{1}(\vk)}+\frac{\mu}{E_{2}(\vk)}
\big ]
-2
,
\label{equ:n-constraint}
\ee
and
\be
0=\bra b^\dagger \cdot\tau^y\sigma^y\cdot b^\dagger \ket_{\rm MF}
=\frac{\partial E_{\rm MF}}{\partial\onsiteansatz} 
=
\int \widetilde{\dif^2\vk}
\big [
\frac{\rho_{2}(\vk)}{E_{2}(\vk)}-\frac{\rho_{1}(\vk)}{E_{1}(\vk)}
\big ]
,
\label{equ:s-constraint}
\ee
under the thermodynamic limit 
$N_{\rm site}^{-1}\sum_{\vk}\to 
\int\widetilde{\dif^2\vk}\equiv
\int \dif^2 \vk/(8\pi^2/\sqrt{3})$, 
where the integral is over the Brillouin zone(BZ) with area $8\pi^2/\sqrt{3}$. 
Note that the $Q_8$ IGG guarantees that the boson numbers on the two orbitals
are the same, so only the total density constraint is needed. 

\subsection{Boson Condensation and Spin Nematic Order}
\label{ssec:condense}

Similar to the standard Schwinger boson mean-field theories\cite{Read-Sachdev-large-N,Sachdev-Read-large-N,Sachdev-kagome}, 
the ordered state can be studied by 
relaxing the total density constraint \eq(\ref{equ:n-constraint}),
and driving the transition to ordered state
by increasing boson density. 
The minima of mean-field dispersions \eq(\ref{equ:dispersion}) are always at 
BZ corners $\pm \vK$, defined by $\vK\cdot \ve_{2,3}=-2\pi/3$. 
So boson condensation will produce a three-sublattice order. 

\begin{figure}
\includegraphics[scale=0.8]{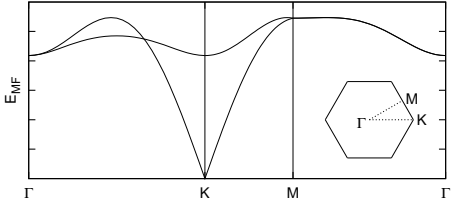}
\caption{
Mean-field dispersions \eq(\ref{equ:dispersion}) along high symmetry lines
at the critical point $\onsiteansatz_c\approx 0.5287$ and
$\mu_c=6-\onsiteansatz_c\approx 5.4713$. 
Inset shows the Brillouin zone and high symmetry points and lines.
Note that at $K$ point there is a critical mode ($\rotor$ in text) 
and a gapped mode ($\gappedrotor$ in text).
}
\label{fig:dispersion}
\end{figure}

Numerical solution of the mean-field critical point 
gives the critical ansatz parameters 
$\onsiteansatz_c\approx 0.5287$
and $\mu_c=6-\onsiteansatz_c\approx 5.4713$, 
the boson dispersions are illustrated in \fig\ref{fig:dispersion}. 
The critical boson density $n_c\approx 0.4492$ is however very low.
Taken at face value this means the spin-1 system with $n=2$
will be deep in the ordered phase. It is conceivable that 
the fluctuations ignored in the mean-field theory 
and/or farther neighbor couplings
may stablize the spin liquid state. 

The boson condensate $\bra \Psi\ket$ is 
a linear combination of the zero-energy eigenvectors of 
the mean-field Hamiltonian \eq(\ref{equ:HMF-k}) at the critical point.
In real space it is 
\be
\bra \Psi_{\vecr}\ket
=
\begin{pmatrix}
1 & e^{-3\pi/4} c & e^{\pi/4} c & 1 \\
e^{3\pi/4} c & -1 & 1 & e^{3\pi/4} c
\end{pmatrix}^T
\cdot
\begin{pmatrix}
e^{+\im \vK\cdot\vecr} w_1 \\
e^{-\im \vK\cdot\vecr} w_2
\end{pmatrix}
,
\ee
where $w_{1,2}$ are complex coefficient, and
constant $c=\sqrt{2+\sqrt{3}}$. 
Note the eigenvectors at $\pm K$
(two rows in the first matrix) form a time-reversal pair. 
Define
\be
\begin{pmatrix}
z_1 \\ z_2
\end{pmatrix}
=
\begin{pmatrix}
e^{2\im \pi /3} & e^{\im \pi /12} c \\
e^{-\im \pi /12} c & -e^{-2\im \pi /3}
\end{pmatrix}\cdot
\begin{pmatrix} w_1 \\ w_2\end{pmatrix}
\ee
and $SU(2)$ rotor field $\rotor$
\be
\rotor=\begin{pmatrix}
 z_1^\nd & z_2\\
-z_2^*\nd & z_1^*
\end{pmatrix}
.
\ee
The boson condensates $\bra \bfield_{\vecr}\ket$ 
[$\bfield$ is defined in \eq(\ref{equ:define-bfield})] 
on the three sublattices are
\bsea
\left \bra
\bfield_A
\right \ket & = & \rotor\cdot
(-\im\tau^y\cdot\nnbond)^T
,
\label{equ:BAtoZ}
\\
\left \bra
\bfield_B
\right \ket & = & \rotor\cdot
(-\im\tau^y\cdot\nnbond)^T (-\im\tau^y\cdot\nnbond)^T
,
\label{equ:BBtoZ}
\\
\left \bra
\bfield_C
\right \ket & = & \rotor
.
\label{equ:BCtoZ}
\esea
It is easy to see that the total dipole moment is
$\bra \vspin\ket=(1/2)\trace[\rotor^\dagger \vsigma \rotor \idmat]=0$.
The nematic directors 
$\trace[\bra\bfield\ket^\dagger \vsigma \bra\bfield\ket \tau^z]$ 
on $A,B,C$ sublattices are given respectively by,
\be
\begin{split}
&
\vdirector_A=-\trace[\rotor^\dagger \vsigma \rotor \tau^y],
\quad
\vdirector_B=\trace[\rotor^\dagger \vsigma \rotor \tau^x],
\\ &
\vdirector_C=\trace[\rotor^\dagger \vsigma \rotor \tau^z]
,
\end{split}
\ee
which are perpendicular to each other. 
Therefore this is the Tsunetsugu-Arikawa spin nematic state. 
For example with $\rotor= (\idmat-\im\tau^z)/\sqrt{2}$
the three directors are $\hat{x},\hat{y},\hat{z}$ 
on $A,B,C$ sublattices respectively,
which is the state depicted in \fig\ref{fig:lattice}.  
The left $SU(2)$ transformations of $\rotor$ are spin rotations, 
the right $SU(2)$ are sublattice-dependent orbital rotations
[the gauge field is only $U(1)\times U(1)\rtimes Z_2$]. 
The PSG transformation rules of $\rotor$
can be derived from the PSG of lattice bosons 
\eqs(\ref{equ:PSGb1}-\ref{equ:PSGb4}),
\bsea
T_{1,2}&:& \rotor\to \rotor\cdot\frac{1}{2}
(-\idmat-\im\tau^x-\im\tau^y-\im\tau^z)^T
,
\\
\sigma_s&:& \rotor\to \rotor
,
\\
\sigma_d&:& \rotor\to \rotor\cdot \frac{1}{\sqrt{2}}(-\im\tau^y+\im\tau^z)^T
,
\\
\treverse&:& \rotor\to \rotor^*
,
\\
\qi&:& \rotor\to \rotor\cdot(-\im\tau^z)^T
,
\\
\qj&:& \rotor\to \rotor\cdot(-\im\tau^x)^T
,
\\
\qk&:& \rotor\to \rotor\cdot(-\im\tau^y)^T
,
\esea
The symmetry allowed form of the low energy action would be 
($\cdif$ is covariant derivative), 
\be
\begin{split}
&
\int \dif^3 x
\Big\{
\trace[(\cdif_\nu \rotor^\dagger)(\cdif_\nu \rotor)]
+m^2 \trace[\rotor^\dagger \rotor]
\\ &
\quad
+u \trace[\rotor^\dagger \rotor]^2
+\dots
\Big\},
\end{split}
\label{equ:low-energy-action}
\ee
with $SO(4)$ symmetry\cite{Grover-quaternion,XuCK-quaternion}. 
Other aspects of the field theory can be found in 
\refes\onlinecite{Grover-quaternion,XuCK-quaternion}
and will not be repeated here.

\subsection{Prototype Wavefunction in Quantum Limit}
\label{ssec:prototype}

The ``quantum limit'' is achieved by 
relaxing the total boson density constraint \eq(\ref{equ:n-constraint})
and going to the low density limit\cite{Sachdev-kagome,Tchernyshyov}
with $\bra n\ket \ll 1$ and $\mu\gg 1$. 
The mean-field constraint equations 
\eqs(\ref{equ:n-constraint}-\ref{equ:s-constraint})
can be solved in power series of $\mu^{-1}$,
\be
\begin{split}
&
\bra n\ket_{\rm MF}=6\mu^{-2}+\frac{135}{2}\mu^{-4}+O(\mu^{-6}),
\\ &
\onsiteansatz=6\mu^{-2}+81\mu^{-4}+O(\mu^{-6})
.
\end{split}
\ee
By inverting the first equation every quantity can also be expressed in 
terms of $\bra n\ket_{\rm MF}$. 

The mean-field bond amplitudes $\bondamplitude_{rr'}$ in \eq(\ref{equ:bondamplitude})
can also be expanded in power series of $\mu^{-1}$ 
and will decay exponentially as $\mu^{-|r-r'|}$
with respect to the distance $|r-r'|$.
For example, bond amplitudes 
$\bondamplitude_{rr'}$ on the nearest- and second- and third-neighbor bonds
are given by
\bsea
\bondamplitude_{\vecr,\vecr+\ve_1} & = & \nnbond\cdot
[\frac{\mu^{-1}}{4}+\frac{15\mu^{-3}}{16}+O(\mu^{-5})]
,
\\
\bondamplitude_{\vecr,\vecr+\ve_2-\ve_3} & = & \im\tau^y\cdot 
[\frac{3\mu^{-3}}{4}+O(\mu^{-5})]
,
\\
\bondamplitude_{\vecr,\vecr+2\ve_1} & = & -\nnbond^T\cdot 
[\frac{3\mu^{-3}}{8}+O(\mu^{-5})]
,
\esea
and those related trivially by cyclic permutations of $\ve_{1,2,3}$ 
(three-fold rotations),
and by $\bondamplitude_{r'r}^\nd=-\bondamplitude_{rr'}^T$. 

The wavefunction in the extreme quantum limit $\mu\to \infty$ simplifies 
to the extreme ``short-range resonating valence bonds'' state,
with nonzero amplitudes only on nearest-neighbor bonds. 
The overall factor of $\bondamplitude_{rr'}$ does not matter. 
The form of $\bondamplitude_{rr'}$ can be fixed by the $Q_8$ IGG without 
calculation and must be proportional to the mean-field ansatz
$\ansatz_{rr'}=\nnbond$ or $-\nnbond^T$. 
To simplify later discussions a factor $-\im$ is applied to 
$\bondamplitude_{rr'}$, so
\be
\bondamplitude_{rr'}=-\im \ansatz_{rr'}=
\left \{
\begin{array}{ll}
-\im\,\nnbond, & {\rm if\ }\vecr'=\vecr+\ve_{1,2,3}, \\
\im\,\nnbond^T,& {\rm if\ }\vecr'=\vecr-\ve_{1,2,3}, \\
0, & {\rm otherwise,}
\end{array}
\right.
\label{equ:prototype-weight}
\ee
with $\nnbond=(\idmat-\im\tau^x-\im\tau^y+\im\tau^z)/2$.
This together with \eqs(\ref{equ:PMFWF}-\ref{equ:loopweight}) defines a
prototype wavefunction which may describe the quaternion spin liquid
(confinement is also possible). 

The loop weight $W_{\ell}$ in \eq(\ref{equ:loopweight})
becomes 
\be
W_{\ell}=N_L\cdot \trace[(-\im\tau^x)\ansatz_{r_1 r_2}
\dots (-\im\tau^x)\ansatz_{r_Lr_1}].
\label{equ:prototype-loop-weight}
\ee
with nearest-neighbor bonds $<r_1r_2>$,\dots,$<r_Lr_1>$,
and $(3/4)^{L/2}$ factor ignored.
The matrix product inside the trace symbol is 
a $SU(2)$ matrix because every factor belongs to $SU(2)$, 
so the trace must be real. 

In fact the trace can only take three values, $\pm 2$ or $0$. 
The proof is the following. 
Denote the number of bonds with orientation along the loop direction
($\chi_{r_i r_{i+1}}=\nnbond$) by $N_{+}$,
and the number of those opposite to the loop direction
($\chi_{r_i r_{i+1}}=-\nnbond^T$) by $N_{-}$.
Due to the three-sublattice structure $N_{+}+2N_{-}\equiv 0\mod 3$. 
This can be formally proved by assigning $Z_3$ numbers $0,1,2$ to
$A,B,C$ sublattices respectively, 
and noting that travelling along(or against) a bond 
increase this number by unity(or two) modulo three. 
Use the identity $-\nnbond^T=(-\im\tau^x)\nnbond\nnbond$ to replace
the $N_{-}$ factors of $\nnbond^T$ by $2N_{-}$ factors of $\nnbond$, 
the matrix product becomes 
$q_1 \nnbond q_2 \nnbond\dots q_{N_{+}+2N_{-}} \nnbond$,
where the $q$s belong to the quaternion group 
$\{\pm \idmat,\pm\im\tau^x,\pm\im\tau^y,\pm\im\tau^z\}$.
Use the commutation relations,
$\nnbond(\pm\im\tau^x)=(\mp\im\tau^z)\nnbond$, 
$\nnbond(\pm\im\tau^y)=(\pm\im\tau^x)\nnbond$, and 
$\nnbond(\pm\im\tau^z)=(\mp\im\tau^y)\nnbond$,
to move all the $q_i$ factors in front of all $\nnbond$ factors, 
the matrix product becomes $q\cdot \nnbond^{N_{+}+2N_{-}}$ 
where $q$ is some quaternion group element. 
Finally use $\nnbond^3=-\idmat$ and the fact 
that $N_{+}+2N_{-}$ is a multiple of three,
the trace becomes $\trace[\pm q]$ which can only be $0$ 
(if $q$ is not $\pm\idmat$)
or $\pm 2$.

\begin{figure}
\includegraphics[scale=0.4]{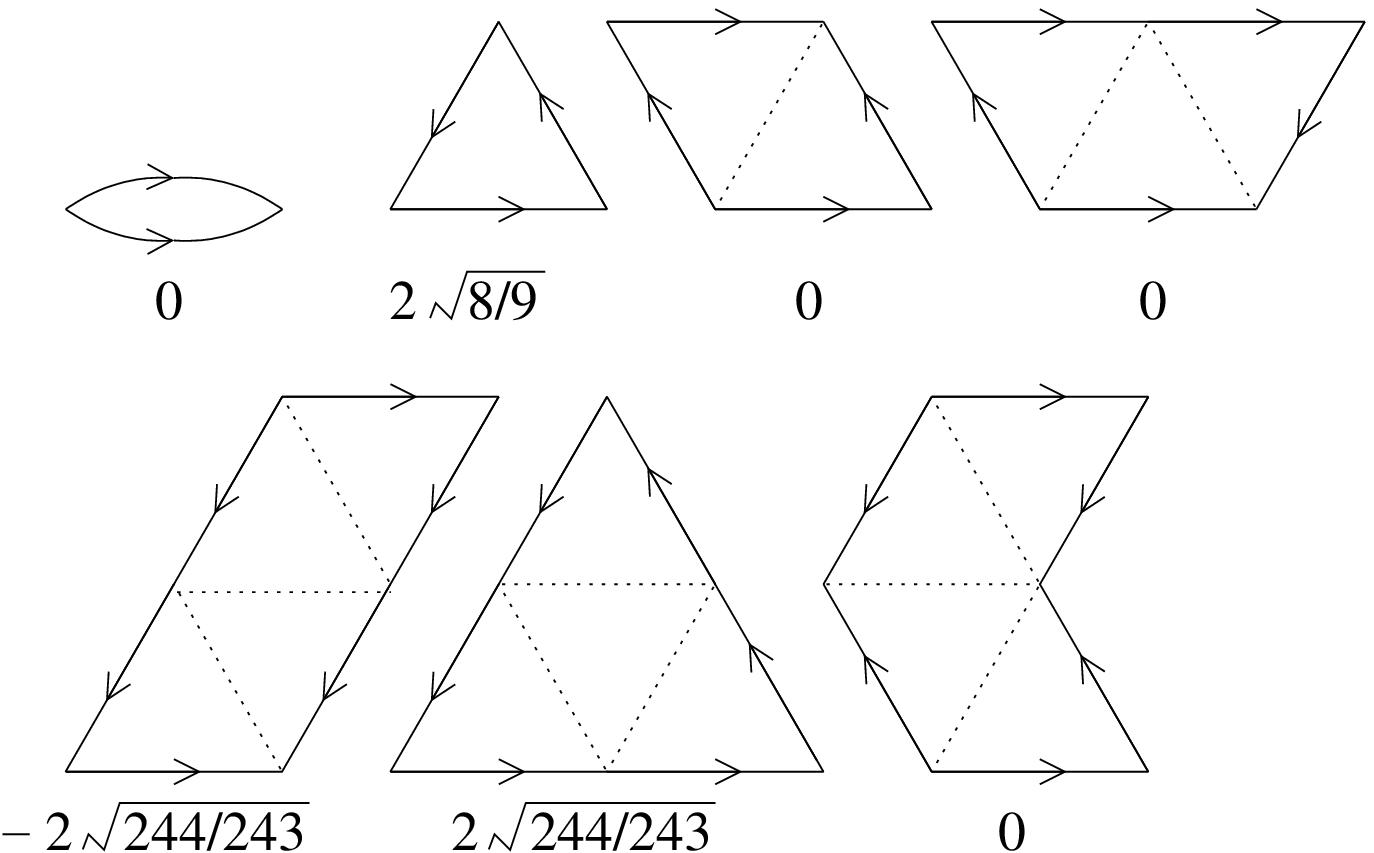}
\caption{
Amplitudes [$W_{\ell}$ in \eq(\ref{equ:loopweight}) without $(3/4)^{L/2}$] of 
several short AKLT loops in the prototype wavefunction given by
\eq(\ref{equ:prototype-weight}). 
The AKLT states reside counter-clockwise on the solid lines. 
Arrow from $r$ to $r'$ means the bond amplitude is 
$
\bondamplitude_{rr'}=-\bondamplitude_{r'r}^T=-\im\,\nnbond
=-\im(\idmat-\im\tau^x-\im\tau^y+\im\tau^z)/2$.
}
\label{fig:loop-weights}
\end{figure}

The AKLT normalization factor $N_L$ approaches unity in the 
long length $L\to \infty$ limit, so may be omitted without changing the 
long distance behavior especially the topological order.
$W_{\ell}$ for several short loops are presented in \fig\ref{fig:loop-weights}. 
Interestingly the weight of ``double-bond'' singlet vanishes, 
so the wavefunction is made purely by AKLT loops of length $L\geq 3$. 

\begin{figure}
\includegraphics[scale=0.6]{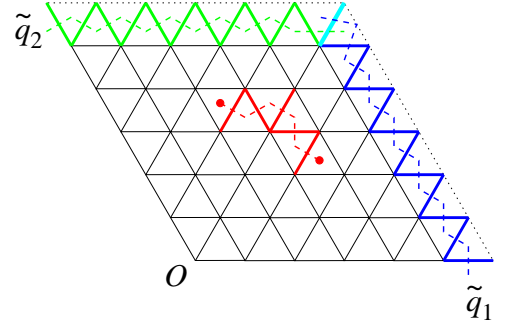}
\caption{(Color online)
$6\times 6$ triangular lattice with periodic boundary condition. 
$O$ indicates the origin site. 
The horizontal green (oblique blue) dash line, 
labelled by $\tilde{q}_2$ ($\tilde{q}_1$) and
cutting through the top row (rightmost column) of 
green (blue) thick bonds, 
defines the creation operator of gauge flux in the 
non-contractible loop along $\ve_2$ ($\ve_1$) direction. 
The red dash line cutting through several red thick bonds 
in the middle defines the creation operator
of a pair of gauge fluxes at the two red dots. 
Gauge flux of class $q \in Q_8$ can be created by setting 
$f_{rr'}\to G_q(r)\cdot f_{rr'}$ for all bonds $rr'$ 
cut by the defining dash line. 
The top-right cyan bond is shared by 
$\tilde{q}_1$ and $\tilde{q}_2$. 
}
\label{fig:gauge-flux}
\end{figure}

The prototype wavefunctions for gauge flux excitations and 
nontrivial topological sectors
can be constructed in this quantum limit by the standard method
(see {e.g.} \refes\onlinecite{Senthil-Fisher,Balents-Fisher-Girvin}). 
Gauge flux on a length-even loop $(r_1\dots r_{2L})$ can be defined as 
$\ansatz_{r_1r_2}^\nd(-\ansatz_{r_2r_3}^*)\dots 
\ansatz_{r_{2L-1}r_{2L}}^\nd(-\ansatz_{r_{2L}r_1}^*)$. 
In the prototype wavefunction,
the flux in each rhombus (unit cell) is $\idmat$. 
For a $6n\times 6m$ lattice ($m,n$ are integers) 
with periodic boundary condition 
the fluxes on the non-contractible(NC) loops are also $\idmat$. 
Creation operators of a pair of local fluxes 
are defined on the string on dual lattice connecting them. 
Creation operators of flux in a NC loop are defined on a NC loop 
of the dual lattice traversing it. 
Creation of gauge flux of class $q\in Q_8$ amounts to 
$\ansatz_{rr'}\to G_q(r)\cdot \ansatz_{rr'}$ 
for all bonds $rr'$ cut by the string or NC loop on dual lattice. 
Examples on a $6\times 6$ lattice are shown in \fig\ref{fig:gauge-flux}. 
Two fluxes on NC loops along $\ve_d$ ($d=1,2$) direction can be 
explicitly defined as
$q_d=
\ansatz_{0,5\ve_d}^\nd(-\ansatz_{5\ve_d,4\ve_d}^*)
\ansatz_{4\ve_d,3\ve_d}^\nd(-\ansatz_{3\ve_d,2\ve_d}^*)
\ansatz_{2\ve_d, \ve_d}^\nd(-\ansatz_{ \ve_d,0}^*)
$.
The 22 topological sectors are given by the conjugacy classes 
of the pair $(q_1,q_2)$ with
the condition $q_1^\nd q_2^{-1} q_1^{-1} q_2^\nd=1$, 
and are explicitly\cite{Grover-quaternion,XuCK-quaternion} 
$(\idmat,\idmat)$, $(\idmat,-\idmat)$, $(-\idmat,\idmat)$, $(-\idmat,-\idmat)$, 
$(\idmat,\im\sigma^a)$, $(-\idmat,\im\sigma^a)$, 
$(\im\sigma^a,\idmat)$, $(\im\sigma^a,-\idmat)$, 
$(\im\sigma^a,\im\sigma^a)$, and 
$(\im\sigma^a,-\im\sigma^a)$, 
with $a=x,y,z$.

\subsection{A Different Perspective}
\label{ssec:different}

Previous discussions are based on uniform 
$\vONdirector_r=\hat{z}$ in constraint \eq(\ref{equ:constraint3}).
A different perspective by allowing non-uniform $\vONdirector_r$ will 
dramatically simplify the picture and results. 
Do a sublattice-dependent orbital rotation
\be
b_{A}\to (\nnbond^T\cdot\im\tau^y)^{-1}\cdot b_{A},\ \,
b_{B}\to (\nnbond^T\cdot\im\tau^y)^{-2}\cdot b_{B},\ \,
b_{C}\to b_{C}
.
\ee
The constraints \eq(\ref{equ:constraint3}) become
\be
\vONdirector_A=\hat{y},\quad
\vONdirector_B=\hat{x},\quad
\vONdirector_C=\hat{z}
.
\ee
The mean-field Hamiltonian \eq(\ref{equ:HMF-triangular}) becomes
\be
\begin{split}
H_{\rm MF}=
\ &
\sum_{\vecr} 
\Big [
\mu\,n_{\vecr}
-(\onsiteansatz\,b^\dagger_{\vecr}\cdot\tau^y\sigma^y\cdot b^\dagger_{\vecr}+h.c.)
\\ &
\phantom{
\sum_{\vecr} 
\Big [
}
-\sum_{d=1}^{3}
(
b^\dagger_{\vecr} \cdot \im\tau^y\otimes\im\sigma^y \cdot b^\dagger_{\vecr+\ve_d}
+{h.c.}
)
\Big ]
,
\end{split}
\label{equ:HMF-triangular-different}
\ee
and is a spin-orbital singlet.
In this gauge choice the quaternion spin liquid state looks like 
``Tsunetsugu-Arikawa orbital nematic state'', 
with $\orbital^y=0,\orbital^x=0,\orbital^z=0$ on 
sublattices $A,B,C$ respectively. 
This ``orbital order'' reduces the orbital $SU(2)$ gauge freedom 
to quaternion group. However it does not break any physical symmetry
in the spin liquid phase. The three-sublattice structure becomes 
physical only upon a spin-orbital-entangled condensation of
$\rotor$ at $\vk=0$, with $\bra \bfield_{\vecr}\ket=\rotor$. 
The PSG of bosons under this gauge is similar to the PSG of $\rotor$. 
In this picture it is clear that 
the low energy theory \eq(\ref{equ:low-energy-action})
contains the coupling of spinon field 
$\spinon\sim\bra\bfield\ket$ to 
the three Higgs fields $\vONdirector_{A,B,C}$, 
\be
\begin{split}
& 
\int\dif^3 x \Big\{
\lambda\,\big(
|\trace[\spinon^\dagger\spinon\vtau]\cdot\vONdirector_A|^2
\\ &
+|\trace[\spinon^\dagger\spinon\vtau]\cdot\vONdirector_B|^2
+|\trace[\spinon^\dagger\spinon\vtau]\cdot\vONdirector_C|^2
\big )
\Big\}
.
\end{split}
\label{equ:coupling-Higgs}
\ee
Note that $\spinon=\rotor+\im\gappedrotor$ where $\gappedrotor$ 
is a different $SU(2)$ rotor field corresponding to the high energy mode
in \fig\ref{fig:dispersion}. This gapped mode has to be included
because $\trace[\spinon^\dagger\spinon\vtau]=
\im \trace[\rotor^\dagger \gappedrotor\vtau]
-\im\trace[\gappedrotor^\dagger \rotor \vtau]$.

The structure of the quantum limit wavefunction is much more 
transparent in this gauge. 
Bond amplitude $\bondamplitude_{rr'}$ in the quantum limit becomes
\be
\bondamplitude_{rr'}=
\left \{
\begin{array}{ll}
\tau^y, & {\rm if\ }\vecr'=\vecr\pm\ve_{1,2,3}, \\
0, & {\rm otherwise.}
\end{array}
\right.
\label{equ:prototype-weight-different}
\ee
The loop weight \eq(\ref{equ:prototype-loop-weight})
becomes
\be
\begin{split}
W_{\ell}
\ &
=N_L\cdot \trace[\tau_{r_1}\tau^y\tau_{r_2}\tau^y
\dots \tau_{r_L}\tau^y]
\\ &
=N_L\cdot \trace[
(-\im \vONdirector_{r_1}^*\cdot\vtau)(-\im \vONdirector_{r_2}^*\cdot\vtau)
\dots (-\im \vONdirector_{r_L}^*\cdot\vtau)]
.
\end{split}
\label{equ:prototype-loop-weight-different}
\ee
$(-\im \vONdirector_{r}^*\cdot\vtau)=-\im\tau^y,-\im\tau^x,-\im\tau^z$
for $r$ on sublattice $A,B,C$ respectively. 
Then it is very easy to see that the trace can only be $\pm 2$ or $0$. 
In fact the trace is nonzero only if 
the number of $A,B,C$ sublattice sites on the loop, $N_{A,B,C}[\ell]$, 
are of the same parity, $N_A[\ell]\equiv N_B[\ell] \equiv N_C[\ell]\mod 2$.

\section{Discussions}
\label{sec:discuss}

In \refe\onlinecite{Grover-quaternion} it was argued that 
no gauge invariant bilinears of the low energy spinon field can be constructed 
to carry spin-1 quantum number. Therefore it was suggested 
the standard projective construction by rewriting spin operators into 
spinon bilinears cannot describe the quaternion spin liquid. 
The argument is indeed true here, and 
there is no gauge invariant spin-1 bilinears of the low energy field $\rotor$. 
However there is a high energy 
branch of spinons $\gappedrotor$ (see \fig\ref{fig:dispersion})
which remains gapped across the spin nematic ordering transition
[see \eq(\ref{equ:coupling-Higgs}) and related discussions].
They together can make gauge invariant spin-1 bilinears 
($\trace[\im\rotor^\dagger \vsigma \gappedrotor]+{c.c}\ \sim$ 
$\vspin$ in \eq(\ref{equ:spin-1})).
This situation was overlooked in 
the analysis of the low energy theory in \refe\onlinecite{Grover-quaternion}. 
So there is no real contradiction. 

An important issue is what spin-1 Hamiltonian may favor 
this quaternion spin liquid as the ground state.
In a numerical study of nearest-neighbor bilinear-biquadratic Heisenberg 
Hamiltonian on triangular lattice\cite{Lauchli-ED},
\be
H=\sum_{<ij>}J\vspin_i\cdot\vspin_j+K(\vspin_i\cdot\vspin_j)^2
,
\ee
it was found that a three-sublattice ``antiferroquadrupolar'' state, 
same as the Tsunetsugu-Arikawa proposal\cite{Tsunetsugu}, 
is the ground state if $K > J > 0$. 
This may serve as the starting point. 
Farther neighbor and mutiple-spin interactions can then be added to 
destroy the long-range order, 
the sign of these terms may be hinted by looking at
the loop-products of spinon pairings  
related to the gauge invariant flux\cite{Wang-SBPSG}. 
For example the term 
$-(b_i^\nd\cdot\ansatz_{ij}^\nd\otimes\im\sigma^y\cdot b_j^\nd)
(b_j^\dagger\cdot\ansatz_{jk}^*\otimes\im\sigma^y\cdot b_k^\dagger)
(b_k^\nd\cdot\ansatz_{kl}^\nd\otimes\im\sigma^y\cdot b_l^\nd)
(b_l^\dagger\cdot\ansatz_{li}^*\otimes\im\sigma^y\cdot b_i^\dagger)
$
defined on a rhombus $ijkl$ 
may favor the quaternion spin liquid. 
After projection to the physical spin-1 space it 
contains a term 
$
-(\vspin_i\cdot\vspin_j)(\vspin_k\cdot\vspin_l)
+(\vspin_i\cdot\vspin_k)(\vspin_l\cdot\vspin_j)
-(\vspin_i\cdot\vspin_l)(\vspin_j\cdot\vspin_k)
-(1/2)(\vspin_i+\vspin_j+\vspin_k+\vspin_l)^2
$
similar to the 4-site ring exchange of spin-1/2, 
but with {\em opposite} sign to that would naturally arise in 
a Hubbard model\cite{ring-exchange}. 

Several possible extensions of the current work exist. 
The two-orbital formalism may also be used to describe 
spin liquids in proximity to other spin nematic or dipole orders. 
One interesting example would be the $Z_4$ 
spin liquid proposed in \refe\onlinecite{XuCK-quaternion}.
A complete classification of PSG in this formalism may be a 
useful guide along this direction. 

The projected spin-1/2 Schwinger boson wavefunctions have been 
numerically studied on small lattices by brute-force evaluation 
of permanents\cite{Motrunich-proj}. 
Generalization to the current case is
likely very hard, because the overlap between a $\spin^z$ basis state and 
the projected wavefunction is not a single but many 
($2^{N_{\rm site}}$) permanents.

The two-orbital AKLT representation can be directly generalized to 
higher spin systems. 
Spin-$S$ can be represented by $2S$ orbitals of spin-1/2 Schwinger bosons, 
with $2S$ single occupancy constraints 
generalizing \eq(\ref{equ:constraint13-z}) on each orbital
and $S(2S-1)$ symmetrization constraints 
generalizing \eq(\ref{equ:constraint2}) between each pair of two orbitals.
The gauge freedom is $[U(1)]^{2S}\rtimes \mathcal{S}_{2S}$ 
where $\mathcal{S}_{2S}$ is the symmetric group of degree $2S$. 
This formalism can describe higher degree multipole orders by boson condensation, 
and spin liquids with even richer gauge structures
(thus richer topological orders)
may be obtained via the projective construction. 

Multiple-orbital fermionic representation
has been considered for general $SU(N)$ spins\cite{Read-Sachdev-SUN},
and used in the context of alkaline-earth cold atom systems\cite{Hermele}. 
Large-$N$ generalization of the multiple-orbital bosonic representation 
may also be useful in theoretical studies.
More recently the two-orbital fermionic representation of spin-1 was 
employed\cite{XuCK-two-orbital-fermion} 
in hope of describing the experimental evidence of 
gapless spin liquid in Ba$_3$NiSb$_2$O$_9$\cite{Balicas}. 

The prototype wavefunction defined by 
\eqs(\ref{equ:PMFWF}-\ref{equ:loopweight},
\ref{equ:prototype-weight}-\ref{equ:prototype-loop-weight},
\ref{equ:prototype-weight-different}-\ref{equ:prototype-loop-weight-different})
may be of some interest by itself. 
It is not clear how to check directly the quaternion structure 
without reference to the mean-field theory.
It is also possible that confinement happens
due to the projection of the mean-field wavefunction.
The confined phase may have nontrivial quantum numbers of
the space group\cite{Read-Sachdev-square}. 
More insight on the
amplitude (matrix trace) are much needed for these purposes. 
And it will be very interesting if the matrix trace form
of the loop amplitudes \eq(\ref{equ:loopweight}) 
can be used to represent other nontrivial phases.

\acknowledgments
The authors thank Todadri Senthil and Tarun Grover for inspiring discussions. 
FW is supported by the MIT Pappalardo Fellowship in Physics. 
CX is supported by the Sloan Research Fellowship.

\end{document}